# Perspectiva intercultural en la formación docente: la Astronomía como punto de entrada.

Intercultural perspective in teacher training: Astronomy as an entry point.

Manente Mayra[1]*, Chadwick Geraldine[2,1], Gangui Alejandro[3].

[1]Instituto de Investigaciones CeFIEC, Facultad de Ciencias Exactas y Naturales, Universidad de Buenos Aires, Int. Güiraldes 2160, CP 1428, Ciudad Autónoma de Buenos Aires, Argentina.

[2]Instituto de Investigaciones en Ciencias de la Educación - CONICET, Universidad de Buenos Aires, Puan 430, Anexo Bonifacio, CP 1406, Ciudad Autónoma de Buenos Aires, Argentina.

[3]Instituto de Astronomía y Física del Espacio - CONICET, Universidad de Buenos Aires, Intendente Güiraldes 2160, CP 1428. Ciudad Autónoma de Buenos Aires, Argentina.

*E-mail: mayra.manente@bue.edu.ar

**Resumen**

Con el objetivo de incorporar en los profesorados de Física una perspectiva intercultural, nos preguntamos qué contenidos disciplinares actuales podrían comenzar a ser abordados desde esta perspectiva. Comenzamos por explorar el estado actual del campo en relación exclusiva a la enseñanza de ciencia intercultural haciendo foco en relaciones directas con la Física, por lo que presentamos aquí una síntesis de las publicaciones de los últimos cinco años encontradas hasta el momento. Las mismas fueron categorizadas en función a su abordaje, como de dimensiones teóricas o de relación con la puesta en práctica. Desarrollamos aquí las siguientes categorías: reflexiones sobre la puesta en práctica, objetos frontera y experiencias de puesta en práctica. Encontramos coincidencias que nos permiten señalar al concepto de esfera celeste como aquel que facilitaría la incorporación de una perspectiva intercultural en los profesorados de Física.

**Palabras clave:** Didáctica de la Astronomía; Formación docente; Educación Científica Intercultural.

**Abstract**

With the aim of incorporating an intercultural perspective into Physics teacher training programs, we ask ourselves which current disciplinary contents could begin to be addressed from this perspective. We started out by exploring the current state of the field exclusively in relation to intercultural science education, focusing on direct relationships with Physics. Therefore, we present here a synthesis of publications from the past five years that we have found so far. These publications were categorized based on their approach, whether related to theoretical dimensions or to the practical implementation. We develop the following categories: reflections on practical implementation, boundary objects, and practical implementation experiences. We find coincidences that allow us to identify the concept of the celestial sphere as the one that would facilitate the incorporation of an intercultural perspective into Physics teacher training programs.

## I. INTRODUCCIÓN

En un mundo globalizado las personas y comunidades de diversos orígenes culturales, antes relativamente lejanas, hoy se encuentran en una muy frecuente y estrecha interacción (UNESCO, 2017). Desde la Organización de las Naciones Unidas para la Educación, la Ciencia y la Cultura, se sostiene que los ciudadanos globalizados deberían contar con competencias interculturales lo suficientemente desarrolladas como para poder llevar a cabo una interacción donde sea posible generar expresiones culturales compartidas mediante el diálogo y una actitud de respeto, en otras palabras, ejercer interculturalidad, lo cual evitaría y revertiría discursos discriminatorios (UNESCO, 2005). En este sentido, consideran a la educación como el principal instrumento internacional que daría respuesta al reto y aseguraría a su vez la educación de calidad para todos, instaurando un conocimiento mutuo, basado en el respeto y el diálogo entre diferentes grupos culturales (UNESCO, 2006). La relación entre educación e interculturalidad ha ido adquiriendo especial relevancia en el campo pedagógico y normativo de nuestro país desde mediados del siglo XX, e incluso se ha intensificado en las últimas décadas. Sin embargo, no ha logrado instaurarse en las escuelas, pudiendo señalar que tal vez una de las causas radique en las reformulaciones parciales que han tenido lugar en la formación docente, las cuales ponen en debate la perspectiva homogeneizante de la escuela moderna pero no logran visibilizar del todo los repertorios culturales (Thisted, 2016).

El último Censo Nacional de Población, Hogares y Viviendas, llevado a cabo en 2022, arrojó que las personas que se reconocen indígenas o descendientes de pueblos indígenas, u originarios, se extienden a lo largo y ancho de todo el país. Dado que nuestra investigación, y su trabajo de campo, se encontrará radicada en la Provincia de Buenos Aires, destacamos que la jurisdicción concentra el 2,1% de la población indígena del total de la Argentina, y que un 21,3% del total de la población residente en viviendas particulares resulta migrante, tanto interno como internacional (INDEC, 2024). Esta diversidad se refleja en las aulas llamando a la reflexión en torno a la supuesta universalidad de los contenidos a enseñar, y como prueba de esto, encontramos que los actuales diseños curriculares de los Profesorados de Educación Secundaria en Física, Química y Biología de la Provincia de Buenos Aires se inscriben en perspectiva intercultural. Presentan como línea de formación a las construcciones políticas, culturales y pedagógicas desde América Latina, lo cual implica asumir y fortalecer la interculturalidad desde el plano de igualdad, reflexionando sobre las prácticas y propiciando un proyecto colectivo no homogeneizador (Subsecretaría de Educación Gobierno de la Provincia de Buenos Aires, 2022). Se espera entonces que los actuales estudiantes de los profesorados del área de ciencias cuenten con las competencias necesarias para llevar a las aulas prácticas pedagógicas interculturales de calidad, poniendo en práctica todo lo aprendido en su etapa de formación. Por ello "*la perspectiva intercultural no puede ser un simple "añadido" al programa de instrucción normal*" (UNESCO, 2006).

Partiendo de este contexto, se definieron las principales preguntas de investigación para la formación doctoral de la primera autora, siendo estas: 1- ¿Qué dispositivos colaborarían en la formación de una perspectiva intercultural para los docentes de física? 2- ¿Cuáles serían los contenidos disciplinares que permitirían comenzar a trabajar con perspectiva intercultural en los profesorados de física?

## II. MARCO TEÓRICO

Convencionalmente dentro de los sistemas educativos tiene lugar la enseñanza de una ciencia que tiende a ser universal, única, eurocéntrica (Walsh, 2005; Krainer y Chaves, 2021; Bonan et al., 2021), donde se priorizan métodos cuantitativos y positivistas, y prevalece una organización en áreas de conocimiento (Delgado y Rist, 2016, en Krainer y Chaves, 2021). Esto puede entenderse como consecuencia de ubicar a la ciencia como uno de los fundamentos centrales de la modernidad, lo cual ha creado *"la ilusión de que el conocimiento es abstracto, desincorporado y deslocalizado llevándonos a pensar que es algo universal, que no tiene casa o cuerpo, ni tampoco género o color"* (Walsh, 2005). Por otra parte, se encuentra la relación entre los pueblos indígenas, la sociedad nacional y el Estado la cual se ha caracterizado históricamente por el conflicto y la tensión, y en dicha relación, las instituciones educativas no permanecen ajenas (Hirsch y Lazzari, 2016). Surge entonces el cuestionamiento respecto a la hegemonía dominante de la ciencia escolar (Chadwick, 2022), el cual puede verse como la consecuencia pedagógica del asumir a los pueblos indígenas y migrantes como sujetos activos capaces de cuestionar su realidad (Mijangos Noh, 2023).

La Interculturalidad entendida como proceso y proyecto intelectual y político, dirigido hacia la construcción de nuevos modos de poder, saber y ser; pretende la transformación de las estructuras e instituciones que responden al orden jerárquico de la lógica colonial (Walsh, 2005). Aparece en escena la Enseñanza de la Ciencia Intercultural (ECI) como aquella que "*propone un modelo dialógico intercultural desde una mirada pluralista, que permita validar los sistemas de conocimiento generados por las comunidades; a través del diálogo se busca transformar identidades y*

*prácticas*" (Riveroll, 2010, en Chadwick, 2018) y que, además, debería caracterizarse por ser *"una educación abierta, flexible, basada en el diálogo intercultural sin admitir asimetrías de ningún tipo"* (Ziradich, 2010). Para lograr que la ECI tenga lugar en las aulas, consideramos que resulta necesario establecer lo que la UNESCO (2017) denomina "objetos frontera", puntos que conservan suficiente significado entre las culturas de modo tal que permiten el diálogo entre las mismas.

## III. METODOLOGÍA

Con el objetivo de conocer el estado actual del campo en relación exclusiva a la ECI, particularmente la ciencia física, se llevó a cabo una selección de publicaciones entre 2018 y 2023, últimos cinco años de estudios sobre el tema. La búsqueda se llevó a cabo en revistas y *journals* del área de educación mayoritariamente de habla hispana, a las cuales se llegaba por medio de buscadores como Scielo o *Google Scholar* utilizando como *keywords* las frases: "educación científica intercultural", "diversidad cultural enseñanza de las ciencias" o "educación intercultural". Si bien la discusión sobre interculturalidad y su repercusión en la educación surge simultáneamente en América Latina, Europa y Estados Unidos de Norteamérica (Abarca-Cariman, 2015, en Acosta Ortiz, Acosta Paz, Díaz, Chadwick y Aduríz-Bravo, 2022), se comenzó por seleccionar trabajos de la región.

Siguiendo la propuesta de Del Rosario Escobar y Buteler (2018) se procedió a crear un código de indexación con el fin de categorizar las publicaciones. Los estudios se clasificaron en dos tipos: Trabajo conceptual (T) y Puesta en práctica (PP). La primera categoría incluye a las publicaciones que estudian el tema desde una perspectiva teórica, reflexiva o crítica de la cuestión actual sin hacer mención a contribuciones didácticas concretas. Esta categoría se conformó por dos subcategorías que se detallan a continuación: (1) Reflexiones teóricas, engloba trabajos cuyo fin es presentar revisiones sobre la epistemología, normativa o sociología de la ECI o EI; (2) Reflexiones sobre la puesta en práctica de ECI, engloba trabajos que ofrecen un análisis o reflexión sobre las políticas educativas de ECI o EI y/o su implementación. La segunda categoría (Puesta en práctica) incluye aquellas publicaciones que presentan resultados de la puesta en práctica de actividades con relación a ECI o EI. Aquí se establecieron dos subcategorías: (1) Objetos frontera, la cual engloba los trabajos cuyo fin es definir puntos en común entre diferentes culturas y (2) Experiencias de puesta en práctica de ECI, engloba prácticas de ECI o EI que se llevaron a cabo en diferentes niveles (primaria, secundaria, superior, capacitación). La Tabla 1 muestra la distribución de los trabajos en función de las categorías y subcategorías.

**Tabla I.** Categorías, subcategorías de análisis y cantidad de trabajos encontrados para cada una de ellas.

| Categorías de análisis | | | | |
|---|---|---|---|---|
| Trabajo conceptual (T) | | Puesta en práctica (PP) | | |
| Reflexiones teóricas (T1) | Reflexiones sobre la puesta en práctica de ECI (T2) | Objetos frontera (PP1) | Experiencias de puesta en práctica de ECI (PP2) | Total |
| 6 | 3 | 4 | 3 | 16 |

## IV. SINTESIS DE LAS PUBLICACIONES

Como se puede apreciar en la Tabla 1, la mayoría de los trabajos (9 de 16) se encontraron en la categoría Trabajo conceptual, y considerando el hecho de que nuestras preguntas de investigación se orientan concretamente a cómo formar docentes con perspectiva intercultural, se dejará el análisis de las reflexiones teóricas para una posterior publicación haciendo foco aquí en el análisis de las categorías: T2, PP1 y PP2.

### A. Reflexiones sobre la puesta en práctica de Enseñanza de la Ciencia Intercultural

Basándose en relevamientos de fuentes secundarias, Corbetta, Bonetti, Bustamante y Vergara Parra (2018) presentan un panorama de la Educación Intercultural Bilingüe y la etnoeducación en América Latina. Respecto a las normativas educativas, los autores señalan que la región se caracteriza por la incorporación de la educación intercultural como una modalidad destinada a los pueblos indígenas. Destacan que el rol del Estado como garante del acceso, permanencia y egreso de los estudiantes aparecería en forma de diversos planes y programas, cuyos únicos destinatarios suelen ser personas indígenas, que en función de su finalidad categorizan en: i) de rescate y uso de las lenguas; ii) de formación de profesionales bilingües y docentes especializados en interculturalidad; iii) de adaptación curricular en el marco de una perspectiva intercultural, y iv) de incorporación de tecnologías a fin de disminuir la brecha existente entre la población indígena y la no indígena. En relación a los niveles educativos a los cuales se destinan dichos programas y planes, se encontró que en la región solo tres países cubren simultáneamente los niveles preescolar, primario y secundario (Argentina, Colombia y Bolivia). Cabe mencionar que Chile cuenta con iniciativas, pero dirigidas únicamente al nivel preescolar. Respecto al nivel superior, señalan que sólo siete países de la región cuentan con programas dirigidos a dicho nivel (Argentina, Bolivia, Brasil, Chile, Colombia, Ecuador y Perú) pero que no se hacen efectivos o se reducen a programas de becas y cupos (Corbetta et al., 2018). Aunque reconocen en la región la presencia de programas de formación e investigación en educación intercultural o intercultural bilingüe, identifican a Bolivia como referente regional, considerando el acceso que ofrece a los pueblos indígenas a la educación superior, sus programas para dicho nivel y la existencia de las universidades como la Universidad Intercultural Indígena Originaria Kawsay, la Universidad Pública de El Alto, la Universidad Indígena Intercultural y sus tres universidades indígenas interculturales: "Tupak Katari", "Casimiro Huanca" y "Apiaguaiki Tumpa".

En relación con cómo se llevan a la práctica la educación intercultural, Corbetta et al. (2018) sostienen la presencia de una inequidad en los logros de aprendizaje de los estudiantes indígenas en América Latina, que se vería reflejada en resultados de evaluaciones estandarizadas y tasas de repitencia. Basándose en el estudio *Inequidad en los logros de aprendizaje entre estudiantes indígenas en América Latina: ¿Qué nos dice TERCE?* de la OREALC-UNESCO, los autores concluyen que para el año 2013 los niños indígenas de sexto grado primaria, de los trece países participantes de la evaluación TERCE (Argentina, Brasil, Chile, Colombia, Costa Rica, Ecuador, Guatemala, Honduras, México, Nicaragua, Panamá, Paraguay y Perú), obtuvieron en promedio una diferencia de -35,6 puntos en la evaluación de ciencias naturales[1], en comparación con el puntaje obtenidos por niños no indígenas. Donde la mayor diferencia se dio en Paraguay con -81,2 puntos, la mínima en Guatemala con 11,3, mientras que en Argentina se obtuvo una diferencia de -25,4. En relación con las tasas de repitencia, el mismo informe señala que para el año 2013 en promedio la tasa de repitencia en la región para los estudiantes de tercer grado primaria era del 16,5 % para los niños no indígenas, número que ascendía a 22,5% en el caso de niños indígenas. Encontrando la mayor diferencia en Paraguay con 25% de tasa de repitencia para niños indígenas y solo un 9% para niños no indígenas, y siendo el caso de Argentina un 25% de tasa de repitencia para niños indígenas y un 15% para niños no indígenas.

Aunque tienen lugar avances en el nivel superior respecto a la incorporación de prácticas de educación intercultural, también se encontró una postura crítica respecto a la formación docente en este aspecto; *"la literatura destaca que uno de los mayores déficits observados en la región se observa en materia de educación superior y formación de especialistas en educación intercultural"* (Corbetta et al., 2018). Este punto resultó ser compartido por Mijangos Noh, quien tras analizar los discursos del Estado de México entre 2000 y 2018 señala que a pesar de los avances en el plano de la política educativa, planes y programas *"la formación de docentes sigue siendo deficiente para atender a la población indígena"* (Mijangos Noh, 2023), y por Espinosa Freire, Herrera Montero y Castellano Gil (2019) quienes durante 2016-2017 observaron 45 actividades docentes de la carrera de Educación Básica en una universidad ecuatoriana. De los resultados estos autores afirman que existe una intencionalidad por parte de los profesores por lograr una formación para el adecuado tratamiento de la dimensión intercultural, pero que se acompaña de un currículum no contextualizado de las asignaturas y metodologías particulares para su tratamiento en el proceso de formación docente no se hacen explícitas.

**B. Respecto a los contenidos conceptuales que actúan como "Objetos frontera"**

De las publicaciones que explicitaban cuáles son los objetos frontera que comúnmente se utilizan para llevar a cabo una ECI, se trabajó con aquellas que estuvieran relacionadas con contenidos de preferencia afines a física y astronomía. El grupo de Investigación en Educación Científica Intercultural (Grupo IECI) del Centro de Formación e Investigación en Enseñanza de las Ciencias de la Facultad de Exactas y Naturales de la Universidad de Buenos Aires, apareció como uno de los referentes a nivel nacional y en los últimos cinco años, han llevado a cabo diferentes investigaciones que permiten identificar puentes conceptuales entre ambas ciencias, particularmente para estudiantes de nivel medio pertenecientes a la cultura qom.

Entendiendo al tiempo como una construcción social cuya concepción varía de cultura a cultura, Chadwick, Bonan y Castorina (2020) realizaron un rastreo a primera instancia de las representaciones vernáculas sobre el tiempo que circulaban en instituciones educativas de nivel secundario del Chaco argentino. Los autores identificaron que las formas de establecer intervalos de tiempo para dicha cultura se basan en los cambios de la naturaleza, los cuales se observan a través de las modificaciones producidas en las plantas, los animales y el canto de las aves. Encontraron también, que los movimientos de algunos grupos de estrellas particulares son indicadores del paso del tiempo en el cielo nocturno relacionados con el ciclo anual. El grupo profundizó puntualmente en las concepciones de la comunidad qom respecto a un cúmulo abierto, o grupo de estrellas sin estructura y asimétrico, visible a ojo desnudo en el cielo nocturno, las Pléyades. Chadwick y Bonan (2018) identificaron cómo se relaciona la aparición de las Pléyades con las estaciones del año y señalan que es mediante la observación nocturna estelar que la comunidad qom logra establecer un calendario basado en sus movimientos vistos desde la Tierra. En ambos trabajos se hace mención de que la observación que realizan los qom de los movimientos del Sol y las Pléyades resultan compatibles con la concepción astronómica conocida como esfera celeste, ya que la primera aparición de estas estrellas en el horizonte oriental (orto helíaco) y su movimiento son indicadores temporales para la comunidad. Otra concepción sobre la cual trabajó el grupo concierne las representaciones sobre los roles de la Luna (*ca´agoxoic*) y sus relaciones con la temporalidad. En Chadwick, Castorina y Bonan (2020), basándose en un relato de la Luna de las madres cuidadoras de la cultura qom titulado Añe nala´, ñe ca´agoxoic qataq aso no´ote (Sol, Luna y No'o te), encontraron una descripción de la Luna como un ser masculino, que muchos identificaron como esposo de Sol, cuyas fases lunares parecen estar relacionadas a su envejecimiento, siendo diferenciadas como: ca'agoxoic (el Luna) recién nacido, creciente, un poco grande, casi completo, recientemente completo, lleno y murió el Luna (ca'agoxoic). Los autores destacan que muchos estudiantes tradujeron la palabra Ca´agoxoic como el equivalente a un mes, lo que lleva a pensar que existe una relación entre los meses y la manera en la que observan el movimiento aparente lunar.

---

[1] La evaluación de ciencias naturales de TERCE abordaba cinco dominios (Salud, Seres vivos, Ambiente, La Tierra y el sistema solar y Materia y energía) y tres procesos cognitivos (Reconocimiento de información y conceptos, Comprensión y aplicación de conceptos y Pensamiento científico y resolución de problemas). El puntaje total era de 92 (UNESCO-OREALC, 2016).

De los resultados mencionados hasta aquí se puede inferir que la observación del cielo y el movimiento de estrellas, Luna, Sol, y posiblemente otros cuerpos, resultan comunes a ambas culturas, pudiendo incluso señalar que "*el concepto de esfera celeste es el principal puente comunicativo entre saberes ancestrales y científicos*" (Chadwick et al., 2020). Esta consiste en una esfera de radio exageradamente grande centrada en los ojos de cada individuo, la cual no deja de ser una construcción mental que surge por la información que reciben los ojos, y varía en función de la ubicación de este.

**C. Sobre las Experiencias de puesta en práctica de la Enseñanza de la Ciencia Intercultural**

En relación con los trabajos que presentaban prácticas de ECI en diferentes niveles educativos (primario, secundario, superior o capacitaciones), fueron pocas las publicaciones encontradas lo cual permite identificar una vacancia dentro del campo. No se encontraron trabajos que mostraran experiencias llevadas a cabo en los niveles obligatorios del sistema educativo abordando conceptos de física o astronomía desde una perspectiva intercultural. En cuanto al nivel superior, encontramos dos publicaciones que mencionan instancias de formación docente en ECI pudiendo señalar que ninguno tuvo lugar al interior de una asignatura, sino que se trataron de instancias de capacitación. Chadwick et al. (2020) llevaron a cabo un taller de 8 horas destinado a los estudiantes de los dos últimos años de un Centro de Investigación y Formación para la Modalidad Aborigen de la Provincia de Chaco. Partiendo del relato sobre el Pájaro Carpintero o Dapichi´[2], propusieron una serie de actividades que invitaban a realizar modificaciones sobre el relato, reflexionar sobre posibles abordajes en las aulas y, a modo de cierre, una actividad metacognitiva. Los autores mencionan que la propuesta fue bien recibida dado que los participantes, todos pertenecientes a comunidades indígenas, mostraron interés en asistir a nuevos talleres. Por su parte, Ithurralde, Becerro, Cordero, Dumrauf y Díaz-Barrios (2023), como parte de una capacitación virtual del Instituto Nacional de Formación Docente, llevada a cabo durante de manera virtual asincrónica durante el 2020, propusieron como parte del encuentro final una instancia de reflexión sobre la enseñanza de las Ciencias Naturales en la escuela primaria desde una perspectiva intercultural, basándose en narrativas indígenas que hacían mención tanto a fenómenos astronómicos, como por ejemplo el solsticio de invierno, como a culturales como el Winoy Tripantu, celebración del pueblo Mapuche que marca el comienzo de un nuevo ciclo en su calendario tradicional. Los autores señalan que en los trabajos finales del curso apreciaron que un 48,7% (94/193) de las entregas presentó algún posicionamiento frente a la diversidad cultural, aunque no detallan qué posicionamientos encontraron.

Ramírez (2023) señala la presencia de preocupaciones y problemas emergentes en la enseñanza de la didáctica general en la Universidad Nacional del Comahue, sede Bariloche, en contexto de heterogeneidad cultural. Proponiendo como solución la reorientación de las nociones didácticas clásicas de modo que atiendan a la complejidad de lo real, con una mirada que revisa las herencias de la didáctica como ciencia moderna. En particular sugiere la concepción ontológica de diseño didáctico sugiriendo la adopción de un diseño didáctico intercultural que

---

[2] Este relato describe al cosmos como una serie de planos o estratos paralelos entre sí que a su vez se estratifican. Los principales estratos son el Cielo (Piguem), la Tierra (Alhua) y el Agua o Submundo (Etaxat), y tienen la característica de estar habitados por seres de poder que transforman su aspecto corporal a medida que los atraviesan

denomina ñiminkaley. Este supone una construcción de conocimiento didáctico en conexión con las características y necesidades locales. La autora no desarrolla cuáles son las preocupaciones y emergentes de la institución o cómo podría un docente materializar el diseño ñiminkaley.

**D. La enseñanza de la astronomía como punto de partida en la formación de docentes de física con perspectiva intercultural**

Encontramos trabajos, pocos, que muestran instancias de formación docente en ECI sosteniendo como punto de entrada conceptos vinculados a la astronomía (Chadwick et al., 2020; Ithurralde et al., 2023) y a su vez, trabajos que identifican como objetos frontera a: la observación del cielo, el movimiento de estrellas, de la Luna, del Sol; y el concepto de esfera celeste (Chadwick y Bonan, 2018; Chadwick et al., 2020). Todos conceptos también vinculados a la astronomía. Desde los primeros años de escolaridad hasta la formación docente, los contenidos de astronomía tienen lugar en las clases de ciencias ya sea, porque están presentes en el diseño curricular del nivel educativo, o bien, por encontrarse en asociación con otros fenómenos físicos. Aun así, estos suelen encontrarse ausentes en las clases de ciencia (Gangui e Iglesias, 2015; Gangui y Adúriz-Bravo, 2017). Esta afirmación no es exclusiva del contexto educativo argentino y tiene diversas causas, pero es sin duda la formación docente una de las principales formas de intervención que puede modificar esta situación sobre la base de acciones sustentadas en la investigación (Amador Rodríguez, Moreno García y Gallego Badillo, 2006).

Las formas de construir conocimiento sobre astronomía no son únicas y singulares, sino que existen diversas maneras de interpretar y resignificar el cielo que compartimos (Chadwick, 2022), apareciendo en escena el concepto de astronomía cultural, la cual por un lado abre un espacio a la reflexión acerca de las formas en que ese conocimiento se construye socioculturalmente (Bastero et al., 2022) y a su vez, valida otros sistemas de conocimiento los cuales no resultan cotidianos en las aulas. Considerando lo planteado hasta aquí, vemos en la interrelación de los conocimientos de la astronomía a enseñar con la astronomía indígena, esencialmente topocéntrica y munida de una visión local de los fenómenos y movimientos celestes tanto nocturnos como diurnos (Gangui, 2011), un punto de partida para comenzar a diseñar dispositivos de formación para los docentes de física, tanto en ejercicio como formación. Destacamos que el modelo didáctico topocéntrico resultará útil, no solo porque describe la mayor parte de los movimientos de los cuerpos que se aprecian desde la superficie terrestre (Gangui e Iglesias, 2015; Galperin et al., 2019; Chadwick y Castorina, 2021), sino porque también quita complejidad de abstracción y resulta más cercano a las ideas previas que podrían traer los estudiantes (Gangui y Adúriz-Bravo, 2017).

## V. CONCLUSIONES

Con el objetivo de conocer el estado actual del campo en relación exclusiva a la ECI, particularmente de la ciencia física, se seleccionaron publicaciones de últimos cinco años. De la síntesis presentada concluimos que la región de América Latina no logra implementar por completo una educación intercultural, lo cual repercute en los estudiantes indígenas ubicándolos en posiciones desfavorables en cuanto a logros de aprendizaje en ciencias respecto a sus pares no indígenas. La formación en perspectiva intercultural de los docentes fue señalada como deficiente a nivel regional, aspecto que en cierto modo se visibiliza en la escasa producción de secuencias didácticas de enseñanza de ciencia intercultural en cualquiera de los niveles educativos.

Los estudios que presentaban experiencias de capacitación en ECI, o bien resultados de investigaciones en relación con su didáctica, coincidieron en utilizar como punto en común entre culturas a conceptos propios de la astronomía, específicamente el concepto de esfera celeste. Es por ello que la pregunta "¿cuáles serían los contenidos disciplinares que permitirían comenzar a trabajar con perspectiva intercultural en los profesorados de Física?" parece parcialmente respondida por la astronomía cultural. Reconociendo que lo presentado aquí es solo una primera aproximación, consideramos como futuros pasos a seguir: profundizar la búsqueda bibliográfica a nivel regional con la intención de recopilar, si las hubiera, experiencias de puestas en práctica en el nivel medio, superior y de capacitación; y ampliar la búsqueda bibliográfica a experiencias de puestas en práctica en países de Europa y los Estados Unidos, dado que también se los menciona como atravesados por la temática.

## AGRADECIMIENTOS

jóvenes Qom sobre el cielo en contextos de Educación Intercultural Bilingüe" del Instituto de Investigaciones en Ciencias de la Educación (IICE), Facultad de Filosofía y Letras, UBA.

**REFERENCIAS**